\documentstyle[12pt,fleqn]{article}

\newcommand{\dip}{\it Dipartimento di Fisica, Universit\`a di Trento,
                      Italia}
\newcommand{\infn}{\it Istituto Nazionale di Fisica Nucleare,
                       Gruppo Collegato di Trento, Italia}

\newcommand{\email}[1]{#1@itncisca.bitnet}
\newcommand{\utf}[1]{\hfill{\sl U.T.F. #1}\par\smallskip\par}
\newcommand{\abs}[1]{\hrule\par\begin{description}\item{Abstract: }
                     \it #1\par\end{description}\hrule\par\bigskip}
\newcommand{\pacs}[1]{\par\bigskip\noindent
                      PACS numbers:\hspace{0.3cm}
                      \parbox[t]{12cm}{\it #1}\par\bigskip}

\newcommand{\ack}[1]{\par\section*{Acknowledgments} #1}
\newcommand{\ca}[1]{{\cal #1}}         
\newcommand{\hs}{\hspace{2cm}}         
\newcommand{\nn}{\nonumber}            
\newcommand{\ap}{\left.}               
\newcommand{\at}{\left(}               
\newcommand{\aq}{\left[}               
\newcommand{\ag}{\left\{}              
\newcommand{\cp}{\right.}              
\newcommand{\ct}{\right)}              
\newcommand{\cq}{\right]}              
\newcommand{\cg}{\right\}}             
\newcommand{\eq}[1]{\begin{equation}#1\end{equation}}  
\newcommand{\eqs}[1]{\begin{eqnarray}#1\end{eqnarray}} 
\newcommand{\ii}{\infty}                         
\newcommand{\X}{\times}                          
\newcommand{\fr}[2]{\mbox{$\frac{#1}{#2}$}}      
\newcommand{\Tr}{\,\mbox{Tr}\,}                  
\renewcommand{\Re}{\,\mbox{Re}\,}                
\newcommand{\lap}{\triangle}                     
\newcommand{\al}{\alpha}

\newcommand{\de}{\delta}
\newcommand{\ep}{\varepsilon}
\newcommand{\ze}{\zeta}

\newcommand{\ka}{\kappa}
\newcommand{\la}{\lambda}
\newcommand{\ro}{\varrho}
\newcommand{\si}{\sigma}

\newcommand{\ph}{\varphi}

\newcommand{\Ga}{\Gamma}

\newcommand{\La}{\Lambda}

\begin{document}

\title{One-loop effective potential on hyperbolic manifolds}

\author{
Guido Cognola$^*$, Klaus Kirsten and Sergio Zerbini\thanks{\infn}
\\
\dip\thanks{Email: \email{cognola}, \email{kirsten}, \email{zerbini}}
}

\date{February 1993}
\maketitle
\utf{284}
\abs{The one-loop effective potential for a scalar field defined on an
ultrastatic space-time whose spatial part is a compact hyperbolic
manifold, is studied using zeta-function regularization for
the one-loop effective action. Other possible regularizations are
discussed in detail. The renormalization group equations are derived
and their connection with the conformal anomaly is pointed out. The
symmetry breaking and the topological mass generation are also
discussed.}


\section{Introduction}

In the last decades, there has been a great deal of investigations
on the properties of interacting quantum field theories in curved
space-time (see for example the text book \cite{birr82b} and
references cited therein).
More recently the inflationary models have been
proposed in order to overcome some difficulties present in the
standard cosmological model (see for example \cite{bran85-57-1} and
references therein).
Several techniques have been employed, among these the background-field
method within path-integral approach
to quantum field theory \cite{jack74-9-1686,toms82-26-2713}, which is very
useful in dealing with the one-loop approximation.
In addition there have been attemps to investigate the role
of the topology and the phenomenon of
the topological mass generation has been discovered
\cite{ford79-70-89,ford80-21-933,toms80-21-2805,toms80-129-334}.
Furthermore  "topological symmetry restoration"
\cite{dena80-58-243,dena80-169-514,kenn81-23-2884}
and later "topological symmetry breaking"
\cite{acto90-7-1463,eliz90-244-387}
have been
investigated in the self-interacting scalar theory.
The non trivial topological space-times which mainly have been
considered are the ones with the torus topology, with the exception
of ref.~\cite{kenn81-23-2884}, in which untwisted
scalar fields on $S^3/\Ga$ are discussed
($\Ga$ is a discrete group of isometries of $S^3$).
Very recently, such manifolds have been considered also by
Dowker et al. \cite{chan92r}.

In this paper we shall consider a self-interacting scalar
field defined on an ultrastatic space-time in which the spatial
section
is a compact hyperbolic 3-dimensional manifold, namely it can be
represented as $H^3/\Ga$, $\Ga$ being a discrete group of isometries of
$H^3$, the Lobachevsky space. It is well known that
such space-times have a  highly non trivial topology and, for a
fixed value of cosmological time, they are spatially homogeneous
isotropic universe models with negative constant curvature.
It is reasonable to assume that such space-times have some physical
relevance. In fact, as it has been stressed in ref.~\cite{elli71-2-7},
they  would affect the occurence of the particle horizons and in
principle they could be tested experimentally by means of observation.
There is also a mathematical motivation based on the Thurston
conjecture \cite{thur82-6-357} which, roughly speaking, states that every
3-dimensional
compact manifold admits a canonical decomposition into eight primitive
pieces. Among such primitive
pieces, the
hyperbolic one is the most important, the other contributions being
exceptional cases. As a consequence, if one is interested in studying matter
fields defined on space-times with a compact spatial section, the compact
hyperbolic case might play an important physical role.
In fact one of the results
achieved in this paper is an expression for the square of the topological
mass. The sign of this depends on the characters of the fields. For
trivial character (untwisted fields) the sign is positive (this might
help to restore symmetry) whereas for twisted fields there could
exist a  mechanism for symmetry breaking.

The plan of the paper is the following.
In Section 2, the scalar model is introduced and
the definition of the
regularized one-loop
effective potential within the background field
method is summarized. It is also shown that the finite part of
the one-loop effective potential in $R^D$ can be determined without
the explicit knowledge of the regularization functions.
In Section 3, making use of the Selberg trace formula,
the one-loop effective potential for $ R\times H^3/\Gamma$ is computed.
In Section 4, the renormalization program is carried out.
In Section 5 a derivation of the renormalization group equations
at one-loop level is presented for the sake of completeness,
paying attention to the role of the conformal anomaly.
In Section 6, symmetry breaking and topological mass generation
are discussed.
The conclusions are summarized in Section 7.
In Appendix A, several regularizations are discussed in order to illustrate
the general result obtained in Section 2.
Finally, Appendix B contains some material on heat  kernel
expansion and zeta-function regularization.

\section{One loop-effective potential}
\label{S:olep}
\renewcommand{\theequation}{\arabic{section}.\arabic{equation}}
\setcounter{equation}{0}

Before we restrict our attention to the hyperbolic manifold
$R\times H^3/\Gamma$, let us deal with a scalar field
$\phi$ defined on an
arbitrary smooth D-dimensional manifold ${\cal M}$
(Euclidean signature), for simplicity without boundary.
To begin with, we recall that the Euclidean generating functional reads

\begin{equation}
W_E[J]=N_E \int d[\phi] e^{-S_E[\phi,J]}
\label{W_E}
\end{equation}
where

\begin{equation}
S_E[\phi,J]=\int_{\cal M}
\aq\fr{1}{2}\phi L \phi +V_o(\phi)-J\phi\cq dvol
\label{S_E}
\end{equation}

\begin{equation}
L=-\lap_D+m^2+\xi R
\label{FopL}
\end{equation}
Here $\lap_D$ is the Laplace-Beltrami operator in ${\cal M}$,
$R$ is the scalar curvature, $\xi$ an arbitrary numerical parameter
and finally $V_o(\phi)$ is the potential describing the
self-interaction of the field.

If we denote  the classical solution by $\phi_c$, the one loop effective
action takes the form

\eqs{
\Gamma[\phi_c]&=&S_E[\phi_c]+\fr{1}{2}\ln\det\fr{A}{\mu^2}=
S_E[\phi_c]+\fr{1}{2}\ln\det\fr{L+V''_o(\phi_c)}{\mu^2}\nn\\
&=&\int\aq V(\phi_c)+\fr{1}{2}Z(\phi_c)
g^{ij}\partial_i\phi_c\partial_j\phi_c+\cdots \cq dvol
\label{EA}
}
in which $A$ is a positive definite elliptic operator of second order
defined on ${\cal M}$ (the small disturbance or fluctuation
operator) and
$\mu$ is an arbitrary normalization parameter
coming from the scalar path-integral measure and necessary in order
to keep the zeta-function dimensionless for all $s$.
The latter equation defines the one-loop effective potential
$V(\phi_c)$. Of course it is equal to the classical one
$V^{(0)}(\phi_c)=V_o(\phi_c)+(m^2+\xi R)\phi_c^2/2$,
plus quantum corrections $V^{(1)}(\phi_c)$ of order $\hbar$.
The quantum corrections to the classical action are formally divergent,
because the eigenvalues of $A$ increase without bound.
One has to use a suitable regularization procedure.

In the following we shall employ a class of regularizations based on the
Schwinger representation (see for example \cite{ball89-182-1}).
In order to work with dimensionless
quantities, we put $B=\mu^{-2} A$.
The regularized determinant of the operator $B$ is defined by

\begin{equation}
(\ln\det B)_\ep= -\int_0^\infty dt\, t^{-1}\ro(\ep,t)
\Tr e^{-tB}
\label{lndetBep}
\end{equation}
where $\ro(\ep,t)$ is a suitable regularization function of the
dimensionless parameter $t$,
which has to satisfy the two requirements we are going to discuss.
First, for fixed $t>0$,
the limit as $\ep$ goes to zero must be equal to one.
Second, for fixed and sufficiently large $\ep$,
$\ro(\ep ,t)$ has to regularize the singularity at $t=0$
coming from the heat kernel expansion

\begin{equation}
\Tr  e^{-tB}\sim\sum_n A_n t^{n-D/2}
\label{TrKt}
\end{equation}
$A_n$ being the integrated Seeley-DeWitt coefficients associated with the
operator $B$. The analytic continuation will be used to reach small
values of $\ep$.
These requirements do not uniquely determine the regularization function
$\ro$. In fact, in the following we shall examine several cases.

Using eq.~(\ref{TrKt}) in eq.~(\ref{lndetBep}) one can easily see that
the number of divergent terms for $\ep\to 0$ is equal to $Q+1$,
$Q$ being the integer part of $D/2$. They are
proportional to the Seeley-De Witt coefficients
$A_0,\dots A_Q$ (which contain the full dependence
on the geometry), the
prefactors depending on the regularization
function \cite{ball89-182-1}.
In the Appendix A, this general result is verified explicitly in
several examples.

In the rest of this Section, in the particular but important
case of $R^{D}$, we would like to show that the
finite part of the effective potential is uniquely
determined, modulo a constant which
can be absorbed by the arbitrary scale parameter $\mu$, and can be
evaluated without making use of the explicit knowledge of the
regularization functions . As usual, in $R^D$ we consider a large
region of volume $V({\cal M})$ and the limit $V({\cal M})\to\ii$
shall be taken at the end of the calculations.
As we will see in the following section, apart from a topological
contribution which will be obtained using the Selberg trace formula,
this calculation is already enough to give the one loop-effective
potential in the constant curvature space-time $R\times H^3/\Gamma$
under consideration.

By definition, the effective potential has to be evaluated
by considering a constant $V''_o$. This means that

\eq{
\frac{\Tr e^{-tB}}{V({\cal M})}
=\frac{\mu^De^{-tf^2/\mu^2}}{(4\pi t)^{D/2}}
}
where $f^2=m^2+V''_o(\phi_c)$ is a positive constant.
As a consequence, from  eqs.~(\ref{EA}) and (\ref{lndetBep}),
we obtain the regularized expression
$V(\ep ,f)$ for $V^{(1)}(\phi_c)$,

\begin{equation}
V(\ep ,f)=\frac{(\ln\det B)_\ep}{2V({\cal M})}
=-\frac{1}{2}\at\frac{\mu^2}{4\pi}\ct^{D/2}
\int_0^\infty dt\,t^{-(1+D/2)}\ro(\ep,t)
e^{-tf^2/\mu^2}
\label{Vep}
\end{equation}

Now it is convenient to distinguish between even ($D=2Q$)
and odd ($D=2Q+1$) dimensions.
In order to derive a more explicit form of the divergent terms for
$\ep \to 0$, let us consider the $Q^{th}$-derivative of $V(\ep )$,
which is

\begin{equation}
\frac{d^QV(\ep ,f)}{dy^Q}
=\frac{(-1)^Q}{2}\at\frac{\mu^2}{4\pi}\ct^{D/2}B_{e,o}(\ep,y)
\label{dQV}
\end{equation}
where

\begin{equation}
B_e(\ep,y)=-\int_0^\infty dt\, t^{-1}\ro(\ep,t)
e^{-ty} = \ln y+b+c_Q(\ep)+\ca{O}(\ep)
\label{Be}
\end{equation}

\begin{equation}
B_o(\ep,y)=-\int_0^\infty dt\, t^{-3/2}\ro(\ep,t)
e^{-ty} = 2\sqrt{\pi y}+b+c_Q(\ep)+\ca{O}(\ep)
\label{Bo}
\end{equation}
In the derivation of the above equations we have made use of the
properties of the regularization functions. Furthermore, in eqs.~(\ref{Be})
and (\ref{Bo}) $b$ is a constant and
$c_Q(\ep)$ is a function of $\ep$, but not of $y$.
We have used the same symbols for even and odd dimensions,
but of course they represent
different quantities in the two cases.

Making use of eqs.~(\ref{Be}) and (\ref{Bo}) in eq.~(\ref{dQV})
a simple integration gives

\begin{equation}
V_{2Q}(\ep,f)=\frac{(-1)^Q}{2Q!(4\pi)^Q}
\aq\ln{\fr{f^2}{\mu^2}}-C_Q+b\cq f^D
+\mu^D\sum_{n=0}^Qc_n(\ep)\at\fr{f}{\mu}\ct^{2n} + \ca{O}(\ep)
\label{V2Q}
\end{equation}

\begin{equation}
V_{2Q+1}(\ep,f)
=\frac{(-1)^Q\Gamma(-Q-1/2)}{2(4\pi)^{Q+1/2}}f^D
+\mu^D\sum_{n=0}^Qc_n(\ep)\at\fr{f}{\mu}\ct^{2n} + \ca{O}(\ep)
\label{V2Q+1}
\end{equation}
where we have set $C_Q=\sum_{n=1}^{Q}\frac{1}{n}$.

The dimensionless integration constants $c_n(\ep)$,
which are divergent for $\ep\to 0$,
define the counter\-terms which must be
introduced in order to remove the divergences.
For the particularly
interesting case $D=4$ one gets

\begin{equation}
V(\ep ,f)=\frac{f^4}{64\pi^2}
\aq\ln\fr{f^2}{\mu^2}-\fr{3}{2}+b\cq
+c_2(\ep)f^4+c_1(\ep)\mu^2f^2+c_0(\ep)\mu^4
\end{equation}
in agreement with the well known
result obtained in refs.~\cite{cole73-7-1888,ilio75-47-165}, where some
specific regularizations were used for $D=4$.

Some remarks are in order. The constants $b$ and $c_n(\ep)$ depend on
the  choice  of the regularization function, but $b$ can
be absorbed by the
arbitrary scale parameter $\mu$.
As a result, the finite part of the effective potential
does not depend on the regularization, as expected.
In the Appendix A, several examples of admissible regularizations  are reported
as an illustration of the above general result.

\section{The effective potential on $R\X H^3/\Ga$}
\setcounter{equation}{0}

In the following we shall consider the scalar field defined
on a topologically non trivial space-time
$R\times H^{D-1}/\Gamma$, that is a space-time
with a hyperbolic $N=D-1$ dimensional manifold as a compact spatial part
and with constant scalar curvature $R=N(N-1)\ka$. Note that $R$ is
also the scalar curvature of $H^N$.
We shall explicitly compute the effective potential only for the physical
interesting case $N=3$, but the extension to any $N$ is
possible.

The manifold $H^N/\Gamma$
can be regarded as a quotient of the Lobachevsky space
$H^N$ by a fixed-point free discontinuous group $\Gamma$ of isometries.
The $\Gamma$ is assumed to be torsion-free.
As usual, for the moment we will consider the case normalized to $\ka
=-1$, so until not otherwise stated all quantities are dimensionless.
The results for arbitrary $\ka$ are simply found by dimensional
reasoning because $H^N/\Gamma$ is a homogeneous manifold.

If $h(r)$ is an even and holomorphic function in a strip of width
$N$ about the real axis, and if $h(r)=\ca{O}(r^{-(N+1+\ep)})$
uniformly in this strip as $r\to\infty$,
then the Selberg trace formula holds
(see for example refs.~\cite{elst87-277-655,selb56-20-47})

\begin{equation}
\sum_{j=0}^{\infty}h(r_j)={V}(\ca{F})\int_0^\infty dr h(r)\Phi_N(r)
+\sum_{\wp}\sum_{n=1}^{\infty}\frac{\chi(P^n(\gamma))}
{S_N(n;l_{\gamma})}\hat h(nl_{\gamma})
\label{STF}
\end{equation}
Here $V(\ca{F})$ is the volume of the fundamental domain $\ca{F}$,
$\hat h$ is the Fourier transform of $h$, $\gamma\in\Gamma$ is an
element of the conjugacy class associated with the length of a
closed geodesic
$l_\gamma$, $\wp$ is a set of primitive closed geodesics on the
compact manifold and each $\gamma\in\wp$ determines
the holonomy element $P(\gamma)$. Moreover,
$S_N(n;l_{\gamma})$ is a known function of the conjugacy class
(see ref.~\cite{chav84b} for details) and $\chi(P)$
represents the characters,
which define all topological inequivalent scalar fields.
The sum on the left hand side of eq.~(\ref{STF}) is over all positive
$r_j$ satisfying

\eq{
r_j^2+\rho_N^2=-\la_j \label{nicoletta}
}
$\la_j$ being the eigenvalues of the
automorphic Laplacian $\lap_N(\Ga,\chi)$ for a group $\Ga$
and representation $\chi$, and $\rho_N=(N-1)/2$.

The quantity $\Phi_N(r)$ is related to the Harish-Chandra
function and reads

\eq{
\Phi_N(r)=\frac{2}{(4\pi)^{N/2}\Ga(N/2)}
\left|\frac{\Ga(\rho_N+ir)}{\Ga(ir)}\right|^2
}
In particular we have
\eq{
\Phi_2(r)=\pi r \tanh \pi r;
\hs \Phi_3(r)=\frac{r^2}{2\pi^2}
}
and the recurrence relation

\eq{
\Phi_{N+2}=\frac{1}{2\pi N}\aq \rho_N^2+r^2\cq\Phi_N(r)
\label{PhiN+2}
}
which permits to obtain any $\Phi_N$ starting from $\Phi_2$ or
$\Phi_3$ according to whether $N$ is even or odd.

We are interested in
\begin{equation}
V(\ep)=-\frac{1}{2V(\ca{M})}
\int_0^\infty dt\, t^{-1}\ro (\ep,t)
\Tr e^{-tB_D}
=-\frac{\mu}{2V(\ca{F})}
\int_0^\infty dt\, t^{-1}\ro (\ep,t)
\frac{\Tr e^{-tB_N}}{(4\pi t)^{1/2}}
\label{VepH3}
\end{equation}
where $B_N=\mu^{-2}\{-\lap_N+m^2+\xi R+V''_o(\phi_c)\}$,
so we choose $h(r)=e^{-t\mu^{-2}(r^2+a^2)}$,
with

\eq{
a^2=\rho_N^2+m^2+\xi R+V''_o(\phi_c)
}
and evaluate $\Tr\exp(-tB_N)$
by using the Selberg trace formula (\ref{STF}).

For $N=3$ we obtain

\eq{
\Tr e^{-tB_3}=
\frac{V(\ca{F})\mu^3e^{-ta^2/\mu^2}}{(4\pi t)^{3/2}}+F_3(t)
\label{TretB3}
}
where $F_3(t)$ is the contribution due to topology.
It can be easily computed for any $N$ and takes the form

\eq{
F_N(t)=\frac{\mu e^{-ta^2/\mu^2}}{(4\pi t)^{1/2}}
\sum_{\wp}\sum_{n=1}^{\infty}
\frac{\chi(P^n(\gamma))}{S_N(n;l_{\gamma})}
e^{-(nl_{\gamma}\mu)^2/4t}
\label{FN}
}

As mentioned in Section 2,
in eq.~(\ref{TretB3}) the contribution due to
the identity element of the group $\Ga$
is formally equivalent to the one we have already considered
in the flat case. The only difference is that $f$ is replaced by $a$.
It has to be noted that this is a peculiarity of 3 dimensions.
When $N=2Q+1>3$, the contribution due to the identity element
of the group $\Ga$ is a polynomial in the variable $t^{-1/2}$,
multiplied by the exponential factor $\exp(-ta^2/\mu^2)$.
It contains $(N-1)/2$ terms proportional to $t^{-3/2}\dots t^{-N/2}$.
Of course, the leading term is formally equivalent to the flat one.
The situation is more complicated for even $N$.
In this case in fact, the identity element of the
group $\Ga$ produces, together with a polynomial of order $N/2$
in the variable $t^{-1}$,
multiplied by the exponential factor $\exp(-ta^2/\mu^2)$,
also a complicated regular function for $t\to 0$.

The function $F_3(t)$ gives a contribution to the one-loop
effective potential, which we refer to as the topological one.
It can be represented in terms of
the
Selberg zeta function $Z(s)$, which may be defined by means of the
equation

\begin{equation}
\frac{d}{ds}\ln Z(s)=\sum_{\wp}\sum_{n=1}^{\infty}
\frac{\chi(P^n(\gamma))}{S_N(n;l_{\gamma})}
e^{-(s-\rho_N)nl_{\gamma}}
\label{SZF}
\end{equation}
It has to be noted that in the evaluation of the topological
contribution $V_T$ to the effective potential,
we can put the regularization function $\ro ( \ep ,t)$ equal to 1
since all the ultraviolet divergences come from the identity part.
This means that the counterterms are not affected by the presence of
the non trivial topology
\cite{toms80-21-928,birr80-22-330,kayb79-20-3052}.
In order to analyze the dependence of the final result on the
curvature, let us now introduce arbitrary constant curvature $R$.
Then we have $l_\gamma \sim 1/\sqrt{|\ka|}$,
$V({\cal F}) \sim |\ka|^{-\frac 3 2}$,
furthermore $a^2=m^2+(\xi -1/6 )R+V''_o(\phi_c)$, and we find
$V(\ep)=V(\ep,a)+V_T$,
$V(\ep,a)$ being given in eq.~(\ref{V2Q}) with $Q=2$ and

\begin{equation}
V_T=-\frac{a^2|\ka|^{-1/2}}{2\pi V(\ca{F})}
\sum_{\wp}\sum_{n=1}^{\infty}\frac{\chi(P^n(\gamma))}
{S_3(n;l_{\gamma})}
\frac{{\bf K}_1(nl_{\gamma}a)}{nl_{\gamma}a}
\label{}
\end{equation}
{}From the latter equation we can see that for $a>0$,
$V_T$ is exponentially
vanishing when $\ka$ goes to zero due to the fact that
$l_{\gamma}$ diverges as $|\ka|^{-1/2}$ when $\ka \to 0$.

Making use of a suitable integral representation for
${\bf K}_1$ \cite{grad80b} and definition (\ref{SZF})
we may write $V_T$ in the final form

\begin{equation}
V_T=-\frac{a^2|\ka|^{-1/2}}{2\pi V(\ca{F})}
\int_1^\infty \;du \;\sqrt{u^2-1}\;
\frac{Z'}{Z}(1+ua|\ka|^{-1/2})
\label{}
\end{equation}
in agreement with the zeta-function calculation \cite{byts92-9-1365}.

\section{Renormalization of $\la\phi^4$ on $R\X H^3/\Ga$}
\setcounter{equation}{0}

In the previous section we have derived the regularized
one-loop effective potential
$V(\phi_c,R)=V^{(0)}(\phi_c,R)+V^{(1)}(\phi_c,R)$ for a scalar field
defined on the manifold $R\X H^3/\Ga$.
We now introduce $R$ into the notation to remind the reader of
the nonvanishing curvature.
The one-loop effective potential
depends on the arbitrary parameter $\mu$, this dependence being
eliminated by the renormalization procedure.
To this aim, here we consider a scalar field self-interacting
by a term of the kind $V_o(\phi)=\la\phi^4/24$ and consider the
classical potential

\eq{
V_c(\phi_c,R))=\La+\frac{\la\phi_c^4}{24}
+\frac{m^2\phi_c^2}{2}+\frac{\xi R\phi_c^2}{2}
+\ep_0R+\frac{\ep_1R^2}{2}
}
$\La$ being the cosmological constant.
Note that one is forced to take into consideration the most general
quadratic gravitational Lagrangian, because the renormalization
procedure gives rise in a natural manner to all such terms
in the curvature \cite{utiy62-3-608}.

We write the renormalized one-loop effective potential in the form

\eq{
V_r(\phi_c,R)=V_c(\phi_c,R)+V_r^{(1)}(\phi_c,R)
}
where
\eq{
V_r^{(1)}(\phi_c,R))=\frac{a^4}{64\pi^2}
\aq\ln\fr{a^2}{\mu^2}-\fr{3}{2}\cq
+V_T(\phi_c,R)+\de V(\phi_c,R)
}
with the counterterm contribution
\eq{
\de V(\phi_c,R)=\de\La+\frac{\de\la\phi_c^4}{24}
+\frac{\de m^2\phi_c^2}{2}+\frac{\de\xi R\phi_c^2}{2}
+\de\ep_0R+\frac{\de\ep_1R^2}{2}
}
reflecting the structure of the divergences.

For convenience now we indicate by $\al=\al_q$ ($q=0,\dots,5$)
the collection of the coupling constants, that is
$(\al_0,\dots,\al_5)\equiv(\La,\la,m^2,\xi,\ep_0,\ep_1)$.
The quantities $\de\al$, which renormalize the coupling constants,
are determined by the following renormalization conditions
\cite{hu84-30-743,berk92-46-1551}

\eq{
\begin{array}{lll}
\La =\ap V_r\right|_{\phi_c=\ph_0,R=0};
&\la  = \ap\frac{\partial^4 V_r}{\partial\phi_c^4}
       \right|_{\phi_c=\ph_1,R=R_1};
&m^2 =\ap\frac{\partial^2 V_r}{\partial\phi_c^2}
       \right|_{\phi_c=0,R=0};\\
\xi  = \ap\frac{\partial^3 V_r}{\partial R\partial\phi_c^2}
       \right|_{\phi_c=\ph_3,R=R_3};
&\ep_0 =\ap\frac{\partial V_r}{\partial R}
       \right|_{\phi_c=0,R=0};
&\ep_1 =\ap\frac{\partial^2 V_r}{\partial R^2}
       \right|_{\phi_c=\ph_5,R=R_5}.
\end{array}
\label{deal}
}
By $\ph_0$ we indicate the value of the field for which the
potential has a true minimum.
For $m^2+\xi R>0$ (respectively $m^2+\xi R <0$)
this value is zero
(respectively $\pm (-6[m^2+\xi
R]/\lambda)^{\frac 1 2}$)
for the classical
potential,
but this may change for the one-loop effective one.
The choice of different values $\phi_c=\ph_q$ and $R=R_q$ in order to
define the physical coupling constants is due to the fact that in
general they are measured at different scales,
the behaviour with respect to a change of scale being determined by
the renormalization group eqs.~(\ref{IRGE}),
which we shall derive in the next Section.
It has to be noted that by a suitable choice of $\ph_q$ and $R_q$,
equations~(\ref{deal}) reduce to the conditions chosen by
other authors (see for example
refs.~\cite{hu84-30-743} and \cite{berk92-46-1551}).

By imposing conditions (\ref{deal})
we obtain the counterterms
(we used an algebraic computing program)

\eqs{
\de\La &=& -\frac{m^2\ph_0^2}{2}-\frac{\la\ph_0^4}{24}
+\frac{\partial^4V_T(\ph_1,R_1)}{\partial\phi_c^4}\frac{\ph_0^4}{24}
+\frac{1}{64\pi^2}\aq M_0^4\at\fr{3}{2}-\ln\fr{M_0^2}{\mu^2}\ct\cp
\nn\\
& &\ap
-\la m^2\ph_0^2\at 1-\ln\fr{m^2}{\mu^2}\ct
+\frac{\la^2\ph_0^4}{12}\at
3\ln\fr{M_1^2}{\mu^2}+\frac{6\la\ph_1^2}{M_1^2}
-\frac{\la^2\ph_1^4}{M_1^4}\ct\cq
\nn\\
\de\la &=&-\frac{\la^2}{32\pi^2}\aq 3\ln\fr{M_1^2}{\mu^2}
+\frac{6\la\ph_1^2}{M_1^2}-\frac{\la^2\ph_1^4}{M_1^4}\cq
-\frac{\partial^4V_T(\ph_1,R_1)}{\partial\phi_c^4}
\nn\\
\de m^2 &=& \frac{\la m^2}{32\pi^2}
\aq 1-\ln\fr{m^2}{\mu^2}\cq
\\
\de\xi &=& -\frac{\la(\xi-1/6)}{32\pi^2}
\aq\ln\fr{M_3^2}{\mu^2}+\fr{\la\ph_3^2}{M_3^2}\cq
-\frac{\partial^3 V_T(\ph_3,R_3)}{\partial\phi_c^2\partial R}
\nn\\
\de\ep_0 &=& \frac{(\xi-1/6)m^2}{32\pi^2}\aq 1-\ln\fr{m^2}{\mu^2}\cq
\nn\\
\de\ep_1 &=& -\frac{(\xi-1/6)^2}{32\pi^2}\ln\fr{M_5^2}{\mu^2}
-\frac{\partial^2 V_T(\ph_5,R_5)}{\partial R^2}
\nn
}
where the fact that the topological contribution
$V_T(\phi_c,R)$ to the effective potential
is exponentially vanishing for $R\to 0$
has been used in writing the latter expressions.
Moreover, we have set

\eq{
M_q^2=a^2(\ph_q,R_q)=m^2+(\xi-\fr{1}{6})R_q+\frac{\la\ph_q^2}{2}
}

The final expression for the renormalized one-loop effective potential
does not depend on the arbitrary scale parameter $\mu$. We have

\eqs{
V_r^{(1)}(\phi_c,R)&=&-\frac{m^2\ph_0^2}{2}-\frac{\la\ph_0^4}{24}
+\frac{\la^2\phi_c^4}{256\pi^2}\aq
\at\ln\fr{a^2}{M_1^2}-\fr{25}{6}\ct+\frac{4m^2(m^2+M_1^2)}{3M_1^4}\cq
\nn\\
& &+\frac{\la\phi_c^2}{64\pi^2}\aq
(\xi-\fr{1}{6})R\at\ln\fr{a^2}{M_3^2}-\fr{3}{2}-\fr{\la\ph_3^2}{M_3^2}\ct
+m^2\at\ln\fr{a^2}{m^2}-\fr{1}{2}\ct\cq
\nn\\
& &+\frac{1}{64\pi^2}
\left[2m^2(\xi-\fr{1}{6})R\at\ln\fr{a^2}{m^2}-\fr{1}{2}\ct
+(\xi-\fr{1}{6})^2R^2\at\ln\fr{a^2}{M_5^2}-\fr{3}{2}\ct\cp
\nn \\ & &\phantom{ \frac{1}{64\pi^2} }
+m^4\ln\fr{a^2}{M_0^2}
-\la m^2\ph_0^2\at\ln\fr{M_0^2}{m^2}-\fr{1}{2}\ct \label{Veffrin}
\\ & &\phantom{ \frac{1}{64\pi^2} }
\ap -\frac{\la^2\ph_0^4}{4}\at\ln\fr{M_0^2}{M_1^2}
-\fr{3}{2}-\fr{4(M_1^2-m^2)(2M_1^2+m^2}{3M_1^4}\ct\cq
\nn\\
& &+V_T (\phi_c ,R)
+\frac{\partial^4V_T(\ph_1,R_1)}{\partial\phi_c^4}
\at\frac{\phi_c^4-\ph_0^4}{24}\ct
\nn\\
& &-\frac{\partial^3 V_T(\ph_3,R_3)}{\partial\phi_c^2\partial R}
\frac{R\phi_c^2}{2}
-\frac{\partial^2 V_T(\ph_5,R_5)}{\partial R^2}\frac{R^2}{2}
\nn
}

Expression (\ref{Veffrin}) notably simplifies in the special but important
case of a conformal invariant scalar field.
First of all we observe that the scalar curvature of a manifold of the kind
$R\X H^N/\Ga$ is just the same as one has for $H^N$. This means that
$R=N(N-1)\ka$ and

\eq{
a^2=m^2+\{\xi-(D-2)/4(D-1)\}R+V''_o(\phi_c)
}
Then we see that for a conformally coupled scalar field,
$a^2=\la\phi_c^2/2$ does not depend on the curvature.
One easily gets

\eqs{
V_r^{(1)}(\phi_c,R)&=&\frac{\la^2\phi_c^4}{256\pi^2}\aq
\ln\fr{\phi_c^2}{\ph_1^2}-\fr{25}{6}\cq
-\frac{\la^2\ph_0^4}{256\pi^2}\aq
\ln\fr{\ph_0^2}{\ph_1^2}-\fr{25}{6}\cq-\frac{\la\ph_0^4}{24}
+V_T(\phi_c ,R)
\nn\\
& &\hspace{\fill}
+\frac{\partial^4V_T(\ph_1,R_1)}{\partial\phi_c^4}
\at\frac{\phi_c^4-\ph_0^4}{24}\ct
-\frac{\partial^3 V_T(\ph_3,R_3)}{\partial\phi_c^2\partial R}
\frac{R\phi_c^2}{2}
\label{Veffconf}
}
which reduces to the potential of Coleman-Weinberg \cite{cole73-7-1888}
in the flat space limit.

\section{The renormalization group equations for $\la\phi^4$}
\setcounter{equation}{0}

In the present section we derive the renormalization group equations
for the $\la\phi^4$ theory \cite{nels82-25-1019} on a 4-dimensional
compact smooth manifold without boundary,
the extension to manifolds with boundary being straightforward.
The derivation is presented only for the sake of completeness and
is valid only at one-loop level. For a more general discussion see for
example \cite{odin92b} and references cited therein.

It is convenient to regularize the one-loop effective action (\ref{EA})
by means of the zeta-function technique
\cite{hawk77-55-133}.
In this way the one-loop effective action is finite and reads

\eq{
\Gamma[\phi_c,g]=S_{c}[\phi_c,g]+\fr{1}{2}\ze'(0;A/\mu^2)
=\int \ca{L}\;\sqrt{g}\;d^4x
\label{regEA}
}
where $\ze'(s;A/\mu^2)$ is the derivative of the zeta-function
related to the small disturbance operator $A=L+\la\phi_c^2/2$.
The classical action is assumed to be of the form

\eq{
S_c[\phi_c,g]=\int\aq\frac{\phi_c L\phi_c}{2}
+\frac{\la\phi_c^4}{24}\cq\sqrt{g}d^4x
+\int\aq\La+\ep_0R+\frac{\ep_1R^2}{2}
+\ep_2C^2\cq\sqrt{g}d^4x
}
$C^2=C^{ijrs}C_{ijrs}$ being the square of the Weyl tensor.
Note however that $C^2=0$ on the manifold under consideration
and so such a term has been omitted in the previous section.

In order to see how the conformal anomaly enters into the game,
we consider a conformal transformation $\tilde g_{ij}=\exp(2\si)g_{ij}$
with $\si$ a constant (scaling).
By the conformal transformation properties of the fields, one can
easily check that $\tilde S_{c}(\al)=S_{c}(\tilde\al)$, where as
before $\al$ represents the collection of all coupling constants and
$\tilde\al$ are all equal to $\al$ apart from $\La$, $m^2$, and $\ep_0$.
For these we have

\eq{
\tilde\La=\La e^{4\si};\hs
\tilde m^2=m^2 e^{2\si};\hs
\tilde\ep_0=\ep_0 e^{2\si}.
}
In the same way, for the eigenvalues $\tilde\mu_n(\al)$
of the small disturbance operator $\tilde A(\al)$ we have
$\tilde\mu_n(\al)=e^{-2\si}\mu_n(\tilde\al)$.
{}From this transformation rule for the eigenvalues,
we immediately get the transformations for
$\ze(s;A/\mu^2)$ and $\ze'(s;A/\mu^2)$. They read

\eqs{
\ze(s;\tilde A/\mu^2)&=&e^{2s\si}\ze(s;A(\tilde\al)/\mu^2)\nn\\
\ze'(s;\tilde A/\mu^2)&=&e^{2s\si}\aq \ze'(s;A(\tilde\al)/\mu^2)
+2\si\ze(s;A(\tilde\al/\mu^2)\cq
\label{}
}
and finally

\eq{
\tilde\Ga(\al)=\Ga(\tilde\al)-\si\ze(0;A(\tilde\al))
\label{}
}

It is well known that $\ze(0;A)$ is related to the
Seeley-DeWitt coefficient $a_{D/2}(x;A)$ by means of the equation

\eq{
\ze(0;A)=\frac{1}{(4\pi)^{D/2}}\int a_{D/2}(x;A)\sqrt{g}d^Dx
\label{bettina}
}
(see the Appendix B for details on zeta-function and heat kernel
expansion).
What is relevant for our case is $a_2(x;A(\tilde\al))$ which is known to
be \cite{dewi65b,gilk75-10-601}

\eqs{
a_2(x;A(\tilde\al))&=&-\frac{\la\lap\phi_c^2}{12}
+\frac{\la\tilde m^2\phi_c^2}{2}
+\frac{\la(\xi-1/6)R\phi_c^2}{2} +\frac{\la^2\phi_c^4}{8}
+\frac{\tilde m^4}{2}\nn\\
&&+\tilde m^2(\xi-\fr{1}{6})R+\frac{(\xi-1/6)^2R^2}{2}
+\frac{C^2}{120}-\frac{G}{360}+\frac{(5-\xi)\lap R}{6}
\label{}
}
where $G=R^{ijrs}R_{ijrs}-4R^{ij}R_{ij}+R^2$ is the Gauss-Bonnet
density, which is a total divergence in 4 dimensions.
Integrating $a_2$ on the manifold and disregarding all total
divergences one finally gets

\eq{
\tilde\Ga(\al)=S_{c}(\tilde\al(\si))+\frac{1}{2}\ze'(0;\tilde A)
=\Ga(\tilde\al(\si))+o(\hbar^2)
\label{}
}
The new parameters $\al(\si)$ are related to the old ones $\al=\al(0)$
by

\eq{
\begin{array}{lll}
\La(\si)-\La = \frac{m^4}{2}\hat\si;
& \la(\si)-\la  = 3\la^2\hat\si;
&m^2(\si)-m^2  = \la m^2\hat\si;
\\
\xi(\si)-\xi  = (\xi-\fr{1}{6})\la\hat\si;
&\ep_0(\si)-\ep_0 = (\xi-\fr{1}{6})m^2\hat\si;
&\ep_1(\si)-\ep_1  = (\xi-\fr{1}{6})^2\hat\si;
\\
\ep_2(\si)-\ep_2  = \frac{1}{30}\hat\si .
&\mbox{}
&\mbox{}
\end{array}
\label{IRGE}
}
Here we have set $\hat\si=-\si/16\pi^2$. With the substitution
$\si\to-2\tau$, the latter equations have been given
for example in refs. \cite{conn83-130-31,hu84-30-743}.
These  important formulae tell us that all parameters
(coupling constants) develop as a result of quantum effects a scale
dependence even if classically they are dimensionless parameters.
This means that in the quantum case we have to define the coupling
constants at some particular scale.
Differentiating eq.~(\ref{IRGE}) with respect to $\si$,
one immediately gets the renormalization group equations
\cite{conn83-130-31,hu84-30-743}.

\section{Symmetry breaking and topological mass generation}
\setcounter{equation}{0}

Let us now concentrate on the physical implications of the one-loop
effective potential, eq.~(\ref{Veffrin}).
In order to analyze the phase
transition of the considered system, let us specialyze to different
cases and to several limits. For this discussion, we will consider
$R_1=R_3=R_5=0$. In this way, for $q=1,3,5$ we have
$M_q^2=m^2+\la\ph_q^2/2$.

Let  us first concentrate on the regime $m^2+\xi R>0$. Then $\phi_c=0$
is a minimum of the classical potential and the expansion of the
quantum corrections around the classical minimum reads
\begin{eqnarray}
V_r &=& \Lambda_{eff}+\frac{(m^2+\xi R+m_T^2)\phi_c^2}{2}
+\frac{\lambda\phi_c^2}{64\pi^2}
\aq m^2\ln\fr{m^2+(\xi- 1/6) R }{m^2}\cp
\nonumber\\
& &\phantom{\Lambda_{eff}}
\ap
+R\at\xi-\fr 1 6\ct
\at\ln\fr{m^2+(\xi-1/6)R}{M_3^2}
-\fr{\lambda\varphi_3^2}{M_3^2}-1\ct\cq
+{\cal O}(\phi_c^4)
\label{guido}
\end{eqnarray}
where $\Lambda_{eff}$ (the effective cosmological constant)
represents a complicated expression
not depending on the background field $\phi_c$ and
$m_T^2$ is the topological contribution
to the mass of the field, which reads

\eq{
m_T^2=
\frac{\sqrt{3}\lambda|R|^{-1/2}}
{4\pi V({\cal F})}
\int_1^{\infty}du\,
(u^2-1)^{-\frac 1 2}
\frac{Z'}{Z}(1+u\sqrt{6m^2|R|^{-1}+1-6\xi})
\label{marina}
}

Let us first restrict our attention to the small curvature limit.
Then, as mentioned, the topological part is negligible and the leading
orders only including the linear curvature terms are easily obtained
from eq.~(\ref{guido}). We obtain

\eq{
V_r =  \Lambda_{eff} +\frac{(m^2+\xi R)\phi_c^2}{2}
-\frac{\la\phi_c^2}{64\pi^2} R(\xi-\fr{1}{6})
\aq\fr{\lambda\varphi_3^2}{M_3^2}-\ln\fr{m^2}{M_3^2}\cq
\label{luciano}
}
This result in fact is true for any smooth manifold, when the constant
curvature $R$ in the considered manifold is replaced by the respective
scalar curvature of the manifold. The recent contribution of
ref.~\cite{berk92-46-1551}
on the Coleman-Weinberg symmetry breaking in a Bianchi type-I universe
is another example of the general structure, which may be proven by
the use of heat-kernel techniques. Due to $R<0$ in our example, for
$\xi<\frac 1 6$ the one loop term will help to break symmetry, whereas
for $\xi > \frac 1 6$ the quantum contribution acts as a positive
mass and helps to restore symmetry. The quantum corrections should be
compared to the classical terms which in general dominate, because the
one loop terms are suppressed by the square of the Planck mass. But if
$\xi$ is small enough the one loop term will be the most important and
will then help to break symmetry.

Let us now consider the opposite limit, that is $|R|\to\infty$ with
the requirement $\xi <0$. In that range the leading order
of the topological part reads

\begin{eqnarray}
m_T^2=\frac{\sqrt{3}\lambda|R|^{-1/2}}{4\pi V({\cal F})}
\int_1^{\infty}\,du\,(u^2-1)^{-\frac 1 2}
\frac{Z'}{Z}(1+u\sqrt{1-6\xi})
+{\cal O}(R^0)\label{monica}
\end{eqnarray}
It is linear in the scalar curvature $R$
because $V({\cal F})\sim |R|^{-\frac 3 2}$ .
The sign of the contribution depends on the choice of the characters
$\chi(P^3(\gamma ))$ ($S_3 (n;l_{\gamma})$ is always positive).
However, for trivial character $\chi =1$ we can say, that the
contribution helps to restore the symmetry, whereas for nontrivial
character it may also serve as a symmetry breaking mechanism.

So, if the symmetry is broken at small curvature,
for $\chi=1$ a symmetry restoration at some critical curvature
$R_c$, depending strongly on the non trivial topology, will take
place.

Let us now say some words on the case $m^2+\xi R <0$,
which includes for example the conformal invariant case.
As we mentioned in section 4 the classical potential has two minima at
$\pm (-6[m^2+\xi R]/\lambda)^{\frac 1 2}$.
So even for $\xi >0$ due to the negative curvature in the given
space-time the classical starting point is a theory with a broken
symmetry. As is seen in the previous discussion, in order to analyze
the influence of the quantum corrections on the symmetry of the
classical potential a knowledge of the function
$Z'(s)/Z(s)$ for values $\Re s< 2$ is required.
Unfortunately, the defining product for the Selberg
zeta-function is only absolutely convergent
in the range $\Re s >2$ \cite{venk79-34-79},
so an analytic continuation of this product
to $\Re s <2$ is needed. Although the analytic continuation exists
(\cite{venk79-34-79}, Theorem 4.1, p.93), using known properties of
this continuation it is not possible to determine some definite
sign of expressions like equation (\ref{marina}). So even for trivial
character $\chi =1$ it is not possible to say, whether the topology
helps to restore the symmetry or not.

\section{Conclusions}

In new inflationary models, the effective cosmological constant is
obtained from an effective potential, which includes quantum
corrections to the classical potential of a scalar field
\cite{cole73-7-1888}. This potential is usually calculated in
Minkowski space-time, whereas to be fully consistent, the effective
potential must be calculated for more general space-times. For that
reason an intensive research has been dedicated to the analysis of the
one-loop effective potential of a self-interacting scalar field in
curved space-time. Especially to be mentioned are
the considerations on the torus
\cite{ford80-21-933,toms80-21-2805,dena80-58-243,acto90-7-1463,eliz90-244-387},
which, however, do not include nonvanishing curvature, and
the quasi-local
approximation scheme developed in \cite{hu84-30-743},
which, however, fails to incorporate global properties of the
space-time.
To overcome this deficiency, we were naturally led to the given
considerations on the space-time manifold $R\times H^3/\Gamma$. Apart
from its physical relevance \cite{elli71-2-7}, this manifold
combines nonvanishing curvature with highly nontrivial topology, still
permitting the exact calculation of the one-loop effective potential
(\ref{Veffrin}) by the use of the Selberg trace formula
(\ref{STF}).
So, at least for $m^2+\xi R >0$ and
trivial line bundles, $\chi=1$, we were able to determine explicitly
the influence of the topology, namely the tendency to
restore the symmetry. Furthermore, this contribution being exponentially damped
for small curvature, we saw, in that regime,
that for $\xi<\frac 1 6$ the
quantum corrections to the classical potential can help to break
symmetry.

\ack{We would like to thank A.~A.~Bytsenko, S.~D.~Odintsov and
L.~Vanzo for helpful discussions and suggestions.
K.~Kirsten is grateful to Prof.~M.~Toller, Prof.~R.~Ferrari and
Prof.~S.~Stringari for the hospitality in the Theoretical Group of the
Department of Physics of the University of Trento.}

\appendix
\section{Appendix: admissible regularizations}
\renewcommand{\theequation}{\Alph{section}.\arabic{equation}}
\setcounter{equation}{0}

For the sake of completeness we give in this appendix a list of admissible
regularization
functions, which are often used in the literature
(see also ref.~\cite{ball89-182-1}). We limit our analysis to $D=4$.

Let us start with
\begin{equation}
\ro_1(\ep,t)=\frac{d}{d\ep}\frac{t^\ep}{\Ga(\ep)}
=\frac{t^{\ep}}{\Gamma(\ep)}\{\ln t-\psi(\ep)\}
\label{18}
\end{equation}
where $\psi(\ep)$ is the logarithmic derivative of $\Ga(\ep)$.
This is the well known zeta-function regularization \cite{hawk77-55-133}.
All requirements are satified. The related $B_e(\ep,y)$ and $V(\ep,f)$, say
$B_1(\ep,y)$ and $V_1(\ep,f)$ read

\begin{equation}
B_1(\ep,y)= y^{-\ep}\ln y=\ln y + \ca{O}(\ep)
\label{19}
\end{equation}

\begin{eqnarray}
V_1(\ep)&=&-\frac{\mu^4y^2}{32\pi^2}
\frac{d}{d\ep}\frac{\Ga(\ep-2)y^{-\ep}}{\Ga(\ep)}
=\frac{\mu^4y^{2-\ep}}{64\pi^2}
\frac{\Gamma(\ep-2)}{\Gamma(\ep)}
\aq\ln y-\psi(\ep-2)+\psi(\ep)\cq\nonumber\\
&=&\frac{f^4}{64 \pi^2}\aq\ln\fr{f^2}{\mu^2}-\fr{3}{2}\cq+\ca{O}(\ep)
\label{20}
\end{eqnarray}
We see that the effective potential is finite,
the divergent terms being removed by
the particular structure of $\ro_1$.

The next regularization we shall consider is closely related to the
above one, and in some sense is associated with the familiar
dimensional regularization in momentum space. It is defined by
\cite{dowk76-13-3224}

\begin{equation}
\ro_2(\ep,t)=t^{\ep}
\label{21}
\end{equation}
{}From this,

\begin{equation}
B_2(\ep,y)= -y^{-\ep} \Gamma(\ep)=
\ln y+\gamma-\frac{1}{\ep} + \ca{O}(\ep)
\label{22}
\end{equation}

\eqs{
V_2(\ep)&=&-\frac{\mu^4y^{2-\ep}}{32\pi^2}
\Gamma(\ep-2)\nn\\
&=&\frac{f^4}{64\pi^2}\aq\ln\fr{f^2}{\mu^2}-\fr{3}{2}+\gamma\cq
-\frac{f^4}{64\pi^2\ep}+\ca{O}(\ep)
\label{23}
}
easily follow, again in agreement with the general result.
Within this regularization, only one divergent term is present.

Another often used regularization is the ultraviolet cut-off
regularization, defined by

\begin{equation}
\ro_3(\ep,t)=\theta(t-\ep)
\label{24}
\end{equation}
In this case we have

\begin{equation}
B_3(\ep,y)=-\Gamma(0,y\ep)
=\ln y+\gamma+\ln\ep + \ca{O}(\ep)
\label{25}
\end{equation}

\eqs{
V_3(\ep)&=&-\frac{\mu^4y^2\Ga(-2,\ep y)}{32\pi^2}\nn\\
&=&\frac{f^4}{64\pi^2}
\aq\ln\fr{f^2}{\mu^2}-\fr{3}{2}+\gamma\cq
-\frac{\mu^4}{64\pi^2\ep^2}+\frac{\mu^2f^2}{32\pi^2\ep}
+\frac{f^4}{64\pi^2}\ln\ep+\ca{O}(\ep)
\label{26}
}
Here we have three divergent terms.
{}From these examples, it is obvious that they depend on
the regularization function.

The fourth regularization reads

\begin{equation}
\ro_4(\ep,t)=e^{-\ep/4t}
\label{27}
\end{equation}
We have

\begin{equation}
B_4(\ep,y)= -2{\bf K}_0(\sqrt{\ep y})
=\ln y+2\gamma-2\ln 2+\ln\ep + \ca{O}(\ep)
\label{28}
\end{equation}

\eqs{
V_4(\ep)&=&-\frac{\mu^4y}{4\ep\pi^2}
{\bf K}_2(\sqrt{\ep y})\nn\\
&=&\frac{f^4}{64\pi^2}
\aq\ln\fr{f^2}{\mu^2}-\fr{3}{2}+2\gamma-2\ln 2\cq
-\frac{\mu^4}{2\pi^2\ep^2}+\frac{\mu^2f^2}{8\pi^2\ep}+
\frac{f^4}{64\pi^2}\ln\ep+\ca{O}(\ep)
\label{29}
}
${\bf K}_\nu$ being the Mc Donald functions.

The last regularization we would like to consider is presented
as an example of the freedom one has.
It is similar to a Pauli-Villars type and
it is defined by ($\al$ being an arbitrary positive constant)

\begin{equation}
\ro_5(\ep,t)=(1-e^{-\al t/\ep})^3
\label{30}
\end{equation}
the power 3 being related to the fact that we are working
in four dimensions. This is a general feature of the
Pauli-Villars regularization.
We have

\begin{equation}
B_5(\ep,y)=\ln\fr{\ep y}{\ep y+3\al}
+3\ln\fr{\ep y+2\al}{\ep y+\al}
=\ln y+\ln\fr{8}{3\al}+\ln\ep+\ca{O}(\ep)
\label{31}
\end{equation}

\eqs{
V_5(\ep)&=&\frac{\mu^4y^2}{64\pi^2}\ag
\aq\ln\fr{\ep y}{\al}-\fr{3}{2}\cq
-3\at 1+\fr{\al}{\ep y}\ct^2\aq
\ln\at 1+\fr{\ep y}{\al}\ct-\fr{3}{2}\cq
\cp\nn\\&&\phantom{\frac{\mu^4y^2}{64\pi^2}}
+\ap 3\at 1+\fr{2\al}{\ep y}\ct^2\aq
\ln\at 2+\fr{\ep y}{\al}\ct-\fr{3}{2}\cq
-\at 1+\fr{3\al}{\ep y}\ct^2\aq
\ln\at 3+\fr{\ep y}{\al}\ct-\fr{3}{2}\cq
\cg\nn\\
&=&\frac{f^4}{64\pi^2}
\aq\ln\fr{f^2}{\mu^2}-\fr{3}{2}+\ln\fr{8\al}{3}\cq
+\frac{3\al^2\ln(16/27)\mu^4}{64\pi^2\ep^2}\\
&&\phantom{\frac{f^4}{64\pi^2}
\aq\ln\fr{f^2}{\mu^2}-\fr{3}{2}+\ln\fr{8\al}{3}\cq}
+\frac{3\al\ln(16/9)\mu^2f^2}{64\pi^2\ep}
+\frac{f^4}{64\pi^2}\ln\ep
+\ca{O}(\ep)\nn
\label{}
}
With this example we conclude the list of possible regularization
functions.

\section{Appendix: heat kernel expansion and zeta-function}
\renewcommand{\theequation}{\Alph{section}.\arabic{equation}}
\setcounter{equation}{0}

For convenience of the reader, here we just report on some results
concerning heat kernel expansion and zeta function.
To start with we consider a second order elliptic differential
operator $A=-\lap+X(x)$ defined on a smooth $D$-dimensional
Riemannian manifold without boundary ($X(x)$ is a scalar function).
The results we are going to discuss have been
quite recently given also for Riemannian manifolds with
boundary
\cite{bran90-15-245,cogn90-241-381,mcav91-8-603,dett92-377-252}
and also in
the presence of torsion \cite{obuk82-108-308,cogn88-214-70}.
The kernel of the operator $\exp(-tA)$, which satisfies a heat type
equation, has an asymptotic expansion, valid for small $t$, given by

\eq{
K_t(x,x;A)\sim\frac{1}{(4\pi t)^{N/2}}
\sum_{n=0}^{\ii}a_n(x;A)t^n
\label{HKE}
}
$a_n(x;A)$ being the spectral (Minakshisundaram, Seeley, DeWitt, Gilkey)
coefficients, which are invariant quantities built up
with curvature and their derivatives.
Some of such coefficients are computed by many
authors (see for example refs.~\cite{bran90-15-245}). In
particular we have (of course $a_0(x;A)=1$)

\eqs{
a_1(x;A)&=&\frac{R}{6}-X \\
a_2(x;A)&=&\fr{1}{2}a_1^2+\fr{1}{6}\lap a_1
+\fr{1}{180}\aq \lap R+R^{ijrs}R_{ijrs}-R^{ij}R_{ij}\cq
}

Sometimes it may be convenient to factorize the exponential $\exp(ta_1)$
and consider an expansion very closely related to eq.~(\ref{HKE}),
that is \cite{jack85-31-2439}

\eq{
K_t(x,x;A)\sim\frac{e^{-t(X-R/6)}}{(4\pi t)^{N/2}}
\sum_{n=0}^{\ii}b_n(x;A)t^n
\label{HKE2}
}
with $b_0=1$, $b_1=0$ and more generally

\eqs{
a_n(x;A)&=&\sum_{l=0}^n \frac{b_{n-l}a_1^l}{l!}\\
b_n(x;A)&=&\sum_{l=0}^n \frac{(-1)^l a_{n-l}a_1^l}{l!}
}
In this way, all coefficients $b_n$ do not depend on $a_1$
\cite{jack85-31-2439}.

The connection between heat kernel and zeta-function is realized by
means of the Mellin representation

\eq{
\ze(s;A)=\frac{1}{\Ga(s)}\int_0^\ii t^{s-1}\Tr e^{-tA} dt
}
Using expansion (\ref{HKE}) in this latter expression, we can get the
meromorphic structure of $\ze(s;A)$ as given by Seeley in
ref.~\cite{seel67-10-172}.
If we suppose $-a_1$ to be a positive function, say $a^2$,
we can obtain the zeta-function as a power series of the kind

\eq{
\ze(s;A/\mu^2)\sim\sum_{n\geq 0}
\frac{\mu^{2s}\Ga(s+n-D/2)}{(4\pi)^{D/2}\Ga(s)}
\int_{{\cal M}}b_na^{-2(s+n-D/2)}\;\sqrt{g}\;d^Dx
}
from which eq.~(\ref{bettina}) follows
and for $D=4$ one obtains furthermore

\eq{
\ze'(0;A/\mu^2)\sim\frac{1}{16\pi^2}\int_{{\cal M}}
\aq \fr{3}{4}a^4-a_2\ln\fr{a^2}{\mu^2}+\sum_{n\geq 1}
\frac{(n-1)!b_{n+2}}{a^{2n}}\cq\sqrt{g}\;d^4x
}
concluding the summary of useful results.

\end{document}